\documentclass[twocolumn,10pt,technote]{IEEEtran}

\usepackage{amsmath}
\usepackage{amssymb}
\usepackage{amsthm}
\usepackage{cite}

\usepackage[final]{graphicx}
\usepackage{color}
\usepackage{subfigure}

\usepackage{hyperref}

\title{Behavior in a Shared Resource Game with Cooperative, Greedy, and Vigilante Players}

\author{Christopher Griffin\footnote{
Dr. Griffin is a faculty member at the Applied Research Laboratory and
Department of Mathematics, 
Penn State University, 
University Park, PA 16802, E-mail: griffinch@ieee.org
}
and George Kesidis\footnote{Dr. Kesidis is a faculty member in the
Departments of Electrical Engineering and
Computer Science and Engineering,
Penn State University, 
University Park, PA 16802, E-mail: gik2@psu.edu}}

\begin{document}
\maketitle

\begin{abstract} 
We study a problem of trust in a distributed system in which a common resource is shared by multiple parties. In such naturally information-limited settings, parties abide by a behavioral protocol that leads to fair sharing of the resource. However, greedy players may defect from a cooperative protocol and achieve a greater than fair share of resources, often without significant adverse consequences to themselves. In this paper, we study the role of a few vigilante players who also defect from a cooperative resource-sharing protocol but only in response to perceived greedy behavior. For a simple model of engagement, we demonstrate surprisingly complex dynamics among greedy and vigilante players. We show that the best response function for the greedy-player under our formulation has a jump discontinuity, which leads to conditions under which there is no Nash equilibrium.  To study this property, we formulate an exact representation for the greedy player best response function in the case when there is one greedy player, one vigilante player and $N-2$ cooperative players. We use this formulation to show conditions under which a Nash equilibrium exists. We also illustrate that in the case when there is no Nash equilibrium, then the discrete dynamic system generated from  fictitious play will not converge, but will oscillate indefinitely as a result of the jump discontinuity. The case of multiple vigilante and greedy players is studied numerically. Finally, we explore the relationship between fictitious play and the better response dynamics (gradient descent) and illustrate that this dynamical system can have a fixed point even when the discrete dynamical system arising from fictitious play does not. 
\end{abstract}

\section{Introduction}
In this paper, we study the problem of trust in a distributed system in which a common resource is shared by many parties or players. In such distributed systems, cooperation and trust are required for the fair and efficient use of a common resource by a plurality of parties/players. Often in such naturally information-limited settings, the players abide by a behavioral protocol that leads to fair sharing of resource. However, a greedy player may defect from a cooperative protocol and achieve a greater than fair share of resources, often without significant adverse consequences if any. This problem has a long history, e.g., \cite{Nowak06,OHLN06,Doz96,Boyd92}, and a broad range of applications - e.g., in \cite{CV83}, the problem of efficient cooperation of two processes that a share resource is studied from a control-theoretic perspective. The more general problem of trust and cooperation remains an active area of research in multiple disciplines \cite{SP05,PB12,RC12}. A principle challenge is attribution, and perhaps even detection, of  deviation from cooperative behavior by some greedy players.

Upon detection of greedy behavior (essentially, detection of a breech of trust), all players may defect from cooperative behavior leading to a less efficient uncooperative (anarchistic) equilibrium or possibly deadlock and a ``tragedy of the commons" \cite{Hardin68}.  In this paper, we consider a much more measured response by only a small number of ``vigilante" players that also defect from cooperative play  but only after greedy behavior has been detected.   The intention of such vigilante play is to entice greedy players back to cooperative play by creating a near deadlock situation in which all players suffer. For an ``objective based" model of engagement, we show surprisingly complex behavior among greedy and vigilante players.

Specifically, we assume a shared resource can be accessed by any of $N$ users at any time, but two users cannot access the resource at the same time. Each user $i$ chooses a probability $q_i$ of accessing the resource at any given time. Thus, the probability that user $i$ can access the resource is:
\begin{equation}
T_i(q_1,\dots,q_N) = q_i\prod_{j \neq i}(1-q_j)
\end{equation}
An example of this model is a synchronous, random-access ALOHA local-area communications network \cite{Kuo95}. In this system, users transmit at random and simultaneous communications cause collision, which results in failed communication. Cooperative use of a resource is common in communications systems  in which all users assume that most, if not all, other users adhere to agreed upon protocols of behavior, e.g., Internet protocols like TCP congestion control, even if cooperation is not in their immediate best interest. Various distributed  mechanisms have been implemented to cooperatively desynchronize demand (e.g., TCP, ALOHA, CSMA).  Typically, when congestion is detected, all end-devices are \textit{expected} to slow down their transmission rates and then slowly increase  again hoping to find a fair and efficient equilibrium. However, if some users employ alternative implementations of the prescribed (``by rule") protocols, e.g., ones that slow down less than they should, or even increase their transmission rate in the presence of congestion, the result could be an unfair allocation or even congestion collapse, see, e.g.,  \cite{Cagalj05,Raya06}.  There is a steadily growing literature on communications that analyzes the equilibria of different distributed network resource allocation \textit{games}, e.g., \cite{Jin02a, Basar02, Wicker03, Jin05, Kesidis10-cdc, Chiang06, Lee07, Long07, Jin07, Cui08, Ma09}; these results are relevant to  more general resource sharing problems. The experience with TCP in particular, e.g., \cite{Shenker02}, has shown that developers do create versions of the protocol that depart from the standard, cooperative (by-rule) congestion-avoidance algorithm, like Turbo TCP, but that the great majority of end-hosts employ the standard cooperative protocol. 

Our objective in this paper is to formulate a model that combines the objective functions of greedy players, vigilante players and cooperative players. Cooperative players follow a prescribed (fair) protocol and are not selfish utility maximizers. Greedy players are selfish utility maximizers whose objective is to take-over the resource. A vigilante player prefers to follow a fair resource sharing protocol, but will increase her transmission rate to punish perceived greediness. As a part of this work, we show that the cyclic behavior induced in \cite{KKG13} through fixed rules can result from a discontinuity in the best-response function.

The remainder of this paper is organized as follows: In Section \ref{sec:Preliminaries} we lay out the preliminary formulae used in the remainder of this paper. In Section \ref{sec:Model} we provide details on our model, including greedy and altruistic player utility functions. We analyze a two-player system in Section \ref{sec:TwoPlayer} we explicitly study a simplified two player shared channel model and characterize the jump discontinuity in the best response function of the greedy player and its effect on Nash Equilibria. In Section \ref{sec:MultiPlayer} we study multi-player systems numerically when there are multiple greedy players or multiple vigilante players and compare our results to the results of better-response dynamics. Finally we provide conclusions and future directions in Section \ref{sec:Future}.

\section{Mathematical Preliminaries}\label{sec:Preliminaries}
Let $q \in [0,1]$ be the transmission probability for a cooperative player in our distributed resource game. In a game with $N$ players, $q = 1/N$, the fair allocation of the resource to a cooperative player. Let $g \in [0,1]$ be the resource access probability of the greedy player. Presumably, $g \geq q$ for any fixed $N$. Finally, let $a \in [0,1]$ be the resource access probability of the vigilante player. Presumably, $a \geq q$ for any $N$. The expected resource access probability for a greedy player is:
\begin{equation}
\Theta(g,a) := g(1-a)\left(1-\frac{1}{N}\right)^{N-2},
\end{equation}
with the corresponding expected resource access probability for the vigilante player is:
\begin{equation}
\Phi(g,a) := a(1-g)\left(1-\frac{1}{N}\right)^{N-2} = \Theta(a,g).
\end{equation}
All other players access the resource with probability:
\begin{displaymath}
\frac{1}{N}(1-a)(1-g)\left(1-\frac{1}{N}\right)^{N-3}.
\end{displaymath}
In the absence of knowledge of the vigilante, the greedy player expects $a = 1/N$ and thus would like to maximize $\Theta\left(g,\tfrac{1}{N}\right)$, which can be accomplished by setting $g = 1$ to obtain a resource access probability of:
\begin{equation}
\Theta_0 := \left(1 - \frac{1}{N}\right)^{N-1}.
\end{equation}
In the absence of knowledge of the greedy player, the vigilante player expects $g = 1/N$ and expects a resource access probability of:
\begin{equation}
\Phi_0 := \frac{1}{N}\left(1 - \frac{1}{N}\right) ^{N} 
\end{equation}

\section{Mathematical Model}\label{sec:Model}
Suppose now the vigilante player expects a (single) greedy player. Using an estimate of her resource access probability $\hat{\Phi}$, an estimate can be obtained for $g$ as:
\begin{equation}
\hat{g} := \left[ \frac{a \left( {\frac {N-1}{N}} \right) ^{N-2}-\hat{\Phi}}{a \left( {\frac {N-1}{N}} \right) ^{N-2}}
\right]_0^1.
\label{eqn:hatg}
\end{equation}
The vigilante player now wishes to enforce fairness unilaterally, by modifying her access probability to punish greedy players. However, it is possible the vigilante player is sensitive to her impact on the community e.g., in the case when the greedy player is only \textit{a little greedy}. In this case, the objective function of the vigilante player to be \textit{minimized} can be written as:
\begin{equation}
U_a(g,a;\rho) := \left(\Theta(g,a)-\Phi_0\right)^2 + \rho\left(a-\frac{1}{N}\right)^2.
\end{equation}
Here $\rho$ is a control parameter that adjusts the extent to which the vigilante is willing to sacrifice her principles of good behavior to punish a greedy player. As we will see, this parameter can have a substantial impact on existence of the underlying system equilibria. 

Conversely, the greedy player wishes to maximize his resource access probability and is willing to violate the communal policy of fairness (e.g., $g = 1/N$) to do so. However, the greedy player realizes there may be a vigilante who will punish him for bad behavior and hence may modulate his behavior back toward the communal norm if he detects his expected resource access probability $\hat{\Theta}$ is well below his desired value $\Theta_0$. The greedy player's objective function to be \textit{minimized} can be formulated as:
\begin{equation}
U_g(g,a;\lambda):=\left(\Theta(g,a)-\Theta_0\right)^2 \cdot\left(1+ \lambda\left(g-\frac{1}{N}\right)^2\right).
\label{eqn:Ug}
\end{equation}
Note that as $\Theta(g,a)-\Theta_0$ approaches zero, then for any fixed value $\lambda$, $\lambda\left(g-\tfrac{1}{N}\right)^2$ also approaches zero and the effect of $\left(g-\tfrac{1}{N}\right)^2$ diminishes. Thus a successful greedy player ignores the fact he is not playing fairly, while an unsuccessful greedy player will throttle back his greediness to try to find a better outcome.  We note that the function $U_g$ has three (first order) critical points given by:
\begin{displaymath}
C_g = \left\{\frac{N-1}{N(1-a)}, \frac{N-3a+2}{(1-a)N}\pm 
\sqrt{-\frac{8}{\lambda}+{\frac { \left( a-2+N \right) ^{2}}{{N}^{2} \left( 
a-1 \right) ^{2}}}}
\right\},
\end{displaymath}
while the function $U_a$ has a single critical point given by:
\begin{equation}
\frac{1}{4}\,{\frac {4\,{g}^{2}-g+2\,\rho}{{g}^{2}+\rho}}.
\end{equation}
Throughout the remainder of this paper, we will study the game in which both the greedy and vigilante players are utility \textit{minimizers} whose decisions affect each other. In the sequel, we refer to this game as $\mathcal{G}(U_g,U_a)$.

\section{Analysis of $N$ Player System}\label{sec:TwoPlayer}
The fact that the objective functions are quartic in $g$ and quadratic in $a$ leads to a complex analytical problem for arbitrary $N \geq 2$. We show that the best response function of the Greedy player may have a jump discontinuity and characterize it completely when it does.

\begin{figure}[htbp]
\centering
\subfigure[Nash Equilibrium]{\includegraphics[scale=0.22]{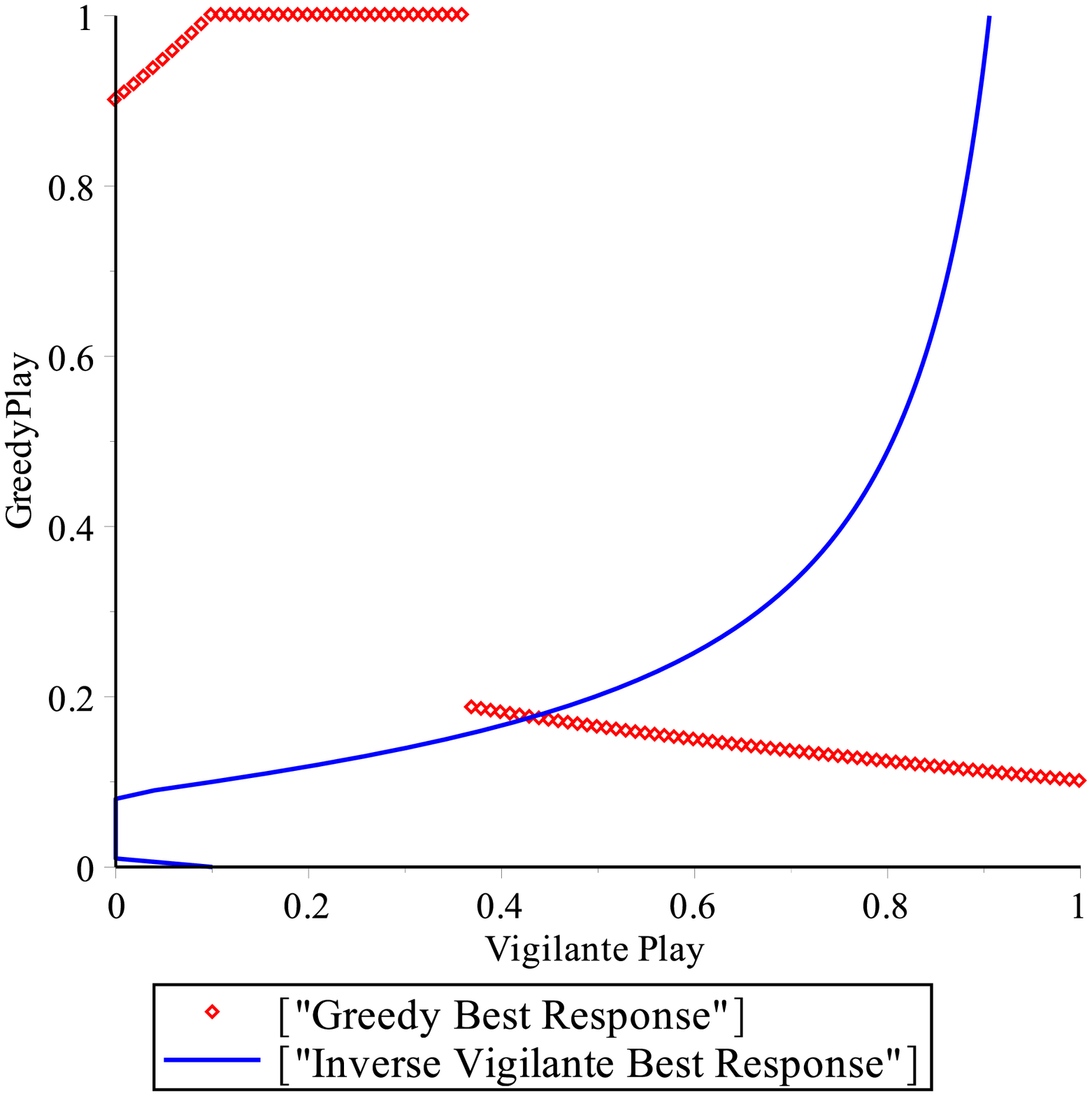}}
\subfigure[No Nash Equilibrium]{\includegraphics[scale=0.22]{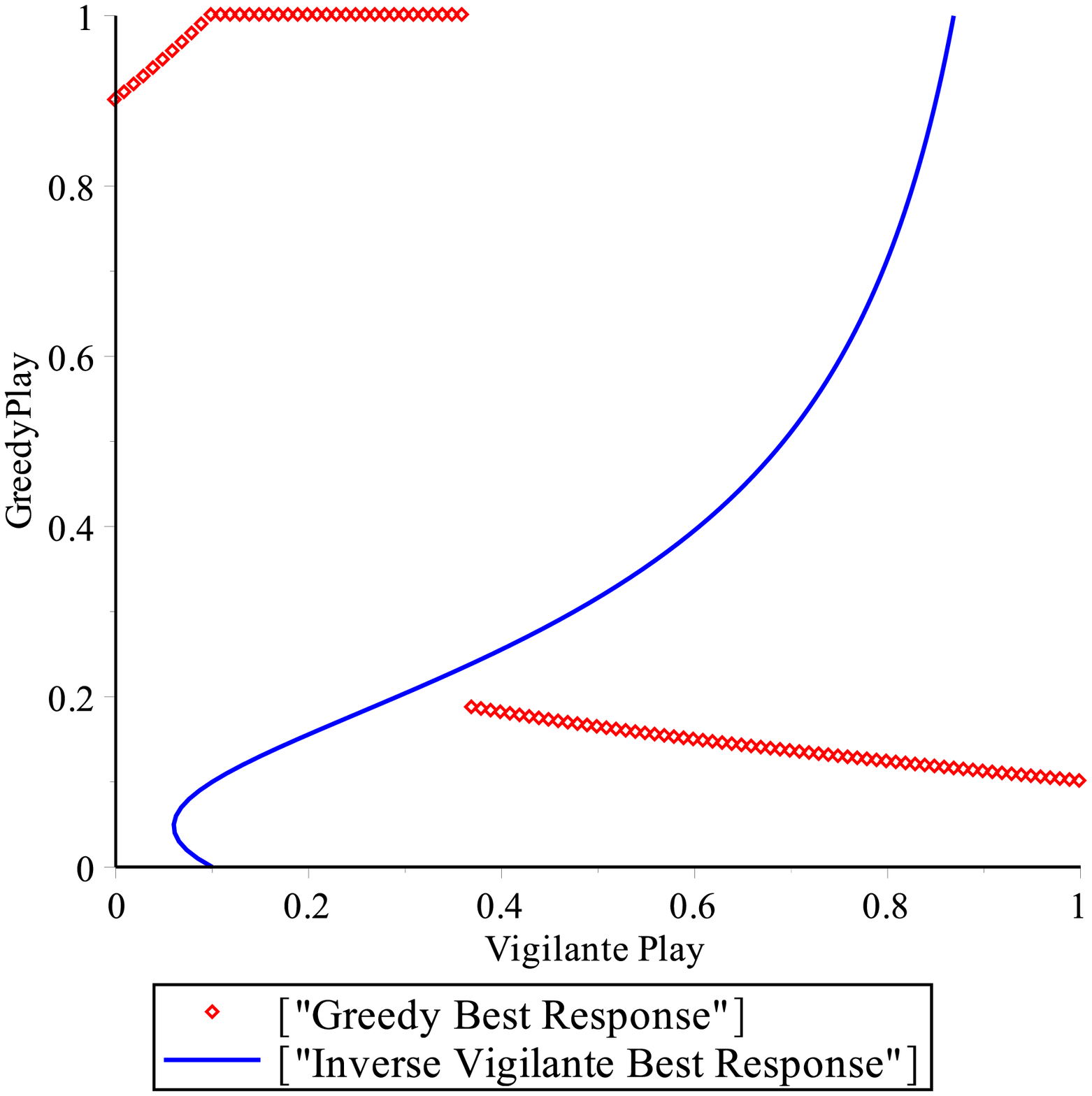}}
\caption{In the first figure, the NE is located at the intersection of the two curves, in this case $\beta_g(a;\lambda)$ and $\beta_a^{-1}(g,\rho)$. In the second figure, no such intersection occurs.}
\label{fig:NEExample}
\end{figure}

Given a value $a \in [0,1]$, the best response function for the greedy player, denoted by $\beta_g(a;\lambda)$ is the set of values of $g$ that minimize $U_g$ for the given value of $a$. We note that when this point-to-set map is a function, then it may be discontinuous, as shown in Figure \ref{fig:NEExample}.
This discontinuity is caused by the non-convexity of $U_g$ in $g$. An interesting result of this phenomenon is the fact that the game $\mathcal{G}(U_g,U_a)$ may not have any Nash equilibrium (NE), leading to interesting discrete time dynamic behavior. 

Let $\beta_a(g;\rho)$ be the best response function for the vigilante player (defined analogously for $\beta_g(a;\lambda)$).  Recall from \cite{Wei95} (Chapter 1) that a pair $(g^*,a^*)$ is a NE if and only if $g^* \in \beta_g(a^*;\lambda)$ and $a^* \in \beta_a(g^*;\rho)$. Suppose that $\beta_a(g;\rho)$ and $\beta_g(a;\lambda)$ are functions (rather than point-to-set maps). A pair $(g^*,a^*)$ is a NE if and only if $g^* = \beta_g(a^*;\lambda)$ and $g^* \in \beta_a^{-1}(a^*;\rho)$ (or likewise $a^* = \beta_a(g^*;\rho)$ and $a^* \in \beta_g^{-1}(g^*;\lambda)$). Here $\beta^{-1}_a$ and $\beta^{-1}_g$ are the usual inverse relations. 

We now illustrate two cases for the game where $N=10$; that is there is one vigilante player and one greedy player and eight cooperative players. In one case, a NE exists and in the other  no NE exists. Fix $\lambda = 10$. For $\rho = 0.001$, a (unique) NE exists while for $\rho=0.01$ there is no NE. The two cases are illustrated in Figure \ref{fig:NEExample}.

We can solve precisely for the point of discontinuity in the best response function and obtain a complete characterization of the discontinuous best-response curve $\beta_g(a;\lambda)$. We have already established that there are three critical points that may come into play in finding (local) minima of the function $U_g$. The discontinuity is caused by the best response moving among two of these three points as well as the boundary value $g = 1$.

We can prove easily that
\begin{equation}
r_1 := \frac{N-1}{N(1-a)}
\end{equation}
is a global minima. To see this, note $U_g(r_1) = 0$ and $U_g$ itself is strictly non-negative and thus $r_1$ must be a global minima since $U_g$ attains $0$ at this value.


We can also see that when
\begin{displaymath}
r_3 := \frac{N-3a+2}{(1-a)N} -  
\sqrt{-\frac{8}{\lambda}+{\frac { \left( a-2+N \right) ^{2}}{{N}^{2} \left( 
a-1 \right) ^{2}}}},
\end{displaymath}
is real and distinct from:
\begin{displaymath}
r_2 := \frac{N-3a+2}{(1-a)N} +  
\sqrt{-\frac{8}{\lambda}+{\frac { \left( a-2+N \right) ^{2}}{{N}^{2} \left( 
a-1 \right) ^{2}}}},
\end{displaymath} 
then it is a local minima. To see this, note that evaluating the second derivative of $U_g$ at $r_3$ yields:
\begin{displaymath}
-\frac{1}{2}\,{N}^{2} \left( {\frac {N-1}{N}} \right) ^{2\,N} \left( s_{{1}}\cdot\gamma +s_{{2}} \right)  \left( N-1 \right) ^{-4}
\end{displaymath}
where:
\begin{displaymath}
s_{{1}}=3\,N\lambda\, \left( -1+a \right)  \left( a-2+N \right)
\end{displaymath}
\begin{multline*}
s_{{2}}=8\,{N}^{2}{a}^{2}-16\,{N}^{2}a-{N}^{2}\lambda-2\,Na\lambda-{a}
^{2}\lambda+8\,{N}^{2}+\\4\,N\lambda+4\,a\lambda-4
\end{multline*}
and
\begin{displaymath}
\gamma=\sqrt {-{\frac {s_{{2}}}{{N}^{2} \left( 1-a \right) ^{2}\lambda}}}
\end{displaymath}
Our assumption that $r_3$ is real implies that $s_2 < 0$. Further, our assumption that $a \in (0,1)$ implies that $s_1 < 0$. Clearly, $\gamma > 0$ (using the customary positive branch of the square root function). It follows that $s_1 \gamma + s_2 < 0$. Thus, $U_g'(r_3) > 0$ and $r_3$ is a local minima. 

As a corollary to the previous result, we note that when it exists and is distinct from $r_3$, the critical point $r_2$ is a local maximum. To see this, we observe that $U_g(g,a;\lambda)$ is a fourth order polynomial in $g$ with a positive coefficient for $g^4$ when we assume $a > 0$ and $\lambda > 0$. The corollary follows from the previous results and this fact.

We now observe that the first critical point $r_1$ is strictly less than 1 when $a < 1/N$. For $a \geq 1/N$, $r_1 \geq 1$. Thus we have proved that for $a \in [0,\tfrac{1}{N}]$, the behavior of $\beta_g(a;\lambda)$ on the left-side of the discontinuity is defined by the function:
\begin{equation}
\beta_g^{-}(a;\lambda):=\min\left\{1,\frac{N-1}{N(1-a)}\right\} = \min\{1,r_1\}.
\end{equation}

Let $a^+$ be the point of discontinuity. We have already shown that $a^+ \geq 1/N$. Clearly now  to the right of $a^+$, the value of $\beta_g(a;\lambda)$ is controlled by the third critical point in $C_g$. Thus we have:
\begin{equation}
\beta_g^{+}(a;\lambda):=
\frac{N-3a+2}{(1-a)N}- 
\sqrt{-\frac{8}{\lambda}+{\frac { \left( a-2+N \right) ^{2}}{{N}^{2} \left( 
a-1 \right) ^{2}}}}
\end{equation}
For $a\in [1/N,a^+]$, $\beta_g(a;\lambda)$ takes on its  boundary value $g^* = 1$. In reality, the best response is a $g^* > 1$, but this is not possible. It now suffices to compute $a^+$. This can be done by solving for the value of $a$ so that:
\begin{equation}
U_g(1,a;\lambda) = U_g\left(r_3,a;\lambda\right)
\label{eqn:Problem}
\end{equation}

Assuming $a^+$ is the (unique) root on $[1/N,1]$ of Equation \ref{eqn:Problem} we now may write:
\begin{equation}
\beta_g(a;\lambda) := \begin{cases}
\beta_g^{-}(a;\lambda) & \text{if $a < a^+$}\\
\beta_g^{+}(a;\lambda) & \text{otherwise}
\end{cases}
\end{equation}

Multiple (non-extraneous) roots for Equation \ref{eqn:Problem}, simply indicate the presence of additional jump discontinuities as the best response moves back and forth between the boundary value $g=1$ and $g = r_3$. In practice we have not observed additional jump discontinuities and we conjecture that for any $\lambda$ there is a unique $a^+ \in [1/N,1]$ that completely characterizes the discontinuity point.

\begin{figure}[htbp]
\centering
\includegraphics[scale=0.25]{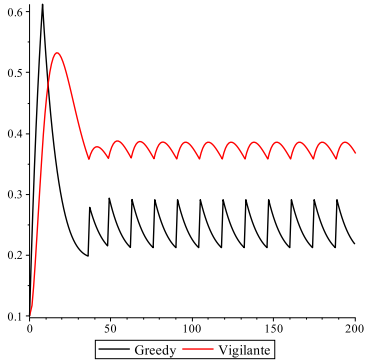}
\caption{The oscillation of the two players strategies in a discrete step iterated game is caused by the jump discontinuity of $\beta_g$.}
\label{fig:Oscillation}
\end{figure}

\begin{figure}[htbp]
\centering
\includegraphics[scale=0.25]{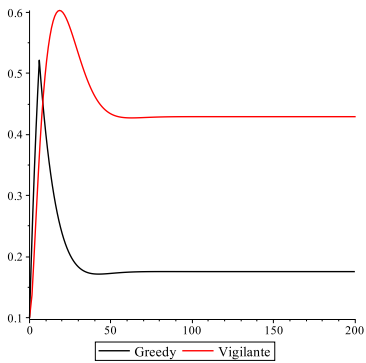}
\caption{The existence of a NE ensures that iterated play converges to a system equilibrium.}
\label{fig:Converge}
\end{figure}

Suppose the Vigilante and Greedy players engage in iterated play and that each player can estimate his/her throughput and hence the other player's strategy. From this information, each player can compute his/her best response using $\beta_g(a;\lambda)$ and $\beta_a(g;\rho)$. The player's strategy at time $t \geq 0$ can then be updated according to the rule:
\begin{gather}
g^{t+1} = (\beta_g(a^t;\lambda) - g^{t})\epsilon_g +  g^t\\
a^{t+1} = (\beta_a(g^t;\lambda) - a^{t})\epsilon_a +  a^t
\end{gather}
Here $\epsilon_g$ and $\epsilon_a$ are parameters that control the extent of the player's jump. In the case when there is no Nash equilibria, we observe oscillatory behavior caused by the jump discontinuity in $\beta_g$. The oscillation size is directly related to the size of $\epsilon_g$ and $\epsilon_a$. This is illustrated in Figure \ref{fig:Oscillation}.
By contrast, when there is a Nash equilibrium, the system converges to it (as would be expected). This is illustrated in Figure \ref{fig:Converge}.

\section{Numerical Analysis of Multi-Player Systems}\label{sec:MultiPlayer}
We now consider two scenarios: (i) We show that the better response behavior given by Jacobi iteration can have convergent behavior, even in the case when there is no Nash equilibrium, illustrating the differences in convergence between better and best response play. (ii) We show that the presence of an additional greedy player yields non-trivial behavioral changes on the part of the greedy and vigilante strategies as a result of the computation of $\hat{g}$ (see Expression \ref{eqn:hatg}).

\subsection{Comparison to Differential Play}
In convex game-theoretic analysis, it is not uncommon to investigate the system of differential equations generated by Jacobi iteration (see e.g., \cite{JK03}). For us, these are defined by:
\begin{equation}
\left\{
\begin{aligned}
\dot{a} = -\frac{\partial U_a(g,a;\rho)}{\partial a}\\
\dot{g} = -\frac{\partial U_g(g,a;\lambda)}{\partial g}
\end{aligned}
\right.
\label{eqn:DiffEQ}
\end{equation}
This model is meant to suggest that the players, rather than computing their best response to (an estimate) of the other player's strategy will follow an (infinitesimal) gradient descent. If a point $(g^*,a^*)$ is an interior NE (that is, it is not on the boundary) then necessarily, ${\partial U_a(g^*,a^*;\rho)}/{\partial a} = {\partial U_g(g^*,a^*;\lambda)}/{\partial g} = 0$; i.e., each interior NE is necessarily a fixed point of the system in Expression \ref{eqn:DiffEQ}. We note that this is a necessary condition for an interior NE, not a sufficient condition in the case of non-convex player objective functions.

\begin{figure}[htbp]
\centering
\includegraphics[scale=0.25]{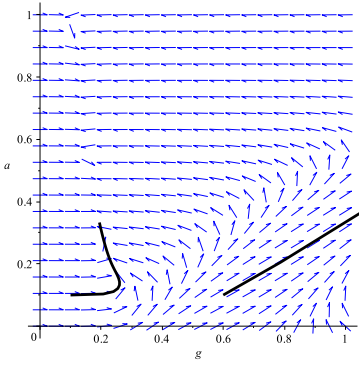}
\caption{The phase portrait of the corresponding Jacobi Iteration with $N=10$, $\lambda = 10$, $\rho=0.01$. Note this attracting fixed is not a NE.}
\label{fig:DiffEQ}
\end{figure}

We have already observed that when $\lambda = 10$ and $\rho = 0.01$, there is no NE. However, there is an interior fixed point for System \ref{eqn:DiffEQ}. Identifying a solution for System \ref{eqn:DiffEQ} requires identifying the roots of a complex set of polynomial equations. These can be solved in closed form (no polynomial has a degree higher than 4) but the closed form solutions do not yield any intuition into the properties of the underlying model. What is interesting, is that there exist real-valued fixed points of the differential equation system for which the system is stable, even when the fixed point is not a NE. In particular, when $\rho = 0.01$, then the point of stability is:
$g \approx 0.203$,
$a \approx 0.297$, 
while for $\rho = 0.001$, the point of stability is
$g \approx 0.175$, 
$a \approx 0.429$, 
where the second fixed point is the same as the Nash equilibrium. The intersection of the best response curves occurs when $\beta_g(a;\lambda) = r_3$ while $\beta_a(g;\rho)$ is (always) computed as:
\begin{equation}
\left[\arg_a\left(\frac{\partial U_a(g,a;\lambda)}{\partial a}=0\right)\right]^1_0
\end{equation}
Thus the intersection of $\beta_g(a;\lambda)$ and $\beta_a(g;\rho)$ must occur at a stability point for System \ref{eqn:DiffEQ}. We can show that in both cases these points are globally stable by analyzing the eigenvalues of the Jacobian matrix of the linearized system. One can verify that when $\rho=0.01$, the eigenvalues of the Jacobian matrix are approximated by $\{-1.501,-0.053\}$, while for $\rho=0.001$ the eigenvalues of the Jacobian matrix are approximated by $\{-1.981,-0.021\}$. Thus by Theorem 3.1 of \cite{Verh06}, the fixed points of the nonlinear systems are stable, even if these points do not correspond to a NE. This is illustrated in Figure \ref{fig:DiffEQ}.
It is also worth noting that this fixed point is not globally attracting. There are initial conditions for which the system moves toward deadlock, which $g=1.0$. These dynamics will only be realized if the players follow a gradient descent strategy, rather than using their best response strategies. 

\subsection{Additional Greedy and Vigilante Players}
An interesting property of this model is its behavior in the presence of multiple greedy or vigilante players. In these cases, it may be impossible for a vigilante player to know the number of greedy players. Consequently, she may choose to assume there is always (exactly) one greedy player and use Expression (\ref{eqn:hatg}) to estimate $g$ for use in $\beta_a(g;\rho)$. In the case when there is more than one greedy player, this will lead the vigilante to overestimate the individual strategies of the greedy players, but this assumption is consistent with what a vigilante could actually communicate. Under this assumption, the vigilante uses the formula:
\begin{equation}
\hat{\Theta}(\hat{g},a):=\hat{g}(1-a)\left(1-\frac{1}{N}\right)^{N-2}.
\end{equation}
Then the vigilante will attempt to minimize:
\begin{equation}
U_a(\hat{g},a;\rho) = \left(\hat{\Theta}(\hat{g},a)-\Phi_0\right)^2 + \rho\left(a-\frac{1}{N}\right)^2.
\label{eqn:Ua}
\end{equation}
Meanwhile, for $M$ greedy players we have:
\begin{equation}
\Theta_i(g_i,g_{-i},a):=g_i(1-a)\left(1-\frac{1}{N}\right)^{N-M-1}\prod_{k\neq i}(1-g_k).
\end{equation} 
The functions $U_{g_i}(g_i,a;\lambda_i)$ are defined analogously. Notice that greedy player $i$ does not need to know about the existence of greedy player $j$ for these objective functions to make sense.

\begin{figure}[htbp]
\centering
\includegraphics[scale=0.25]{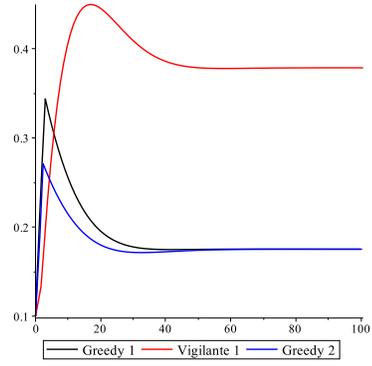}
\caption{Convergence in the case when $N=10$, $\lambda_1 = \lambda_2 = 10$ and $\rho = 0.01$ and there are two greedy players.}
\label{fig:TwoGreedy}
\end{figure}

In the case when there are additional vigilante players, then we modify Expression (\ref{eqn:Ua}) slightly to:
\begin{displaymath}
U_{a_i}(\hat{g},a_i;\rho_i) = \left(\hat{\Theta}(\hat{g},a_i)-\Phi_0\right)^2 + \rho_i\left(a_i-\frac{1}{N}\right)^2
\end{displaymath}
Additional vigilante players will simply see vigilante activity as the result of a greedy play. Some interesting behaviors occur in both the case when there are additional greedy or vigilante players. In the case when $\lambda_1 = \lambda_2 = 10$ and $\rho = 0.01$, we obtain convergence to a NE, unlike when there was only a single greedy player with $\lambda = 10$ and $\rho=0.01$. This is illustrated in Figure \ref{fig:TwoGreedy}.   
In this case, the two greedy player converge to the same value at equilibrium. There are still parameters (as before) for which the system does not converge, but it is interesting to note that the introduction of additional greedy players causes convergence for parameters that were non-convergent in the single greedy-player case.

\begin{figure}[htbp]
\centering
\includegraphics[scale=0.25]{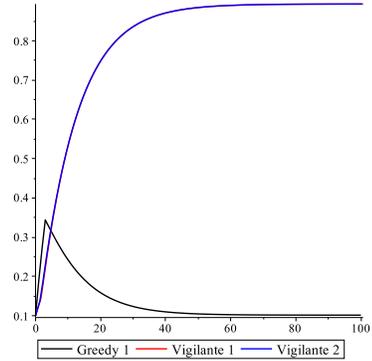}
\caption{Convergence in the case when $N=10$, $\lambda = 10$ and $\rho_1 = \rho_2 = 0.001$ and there are two vigilante players.}
\label{fig:TwoVigilante1}
\end{figure}

\begin{figure}[htbp]
\centering
\includegraphics[scale=0.25]{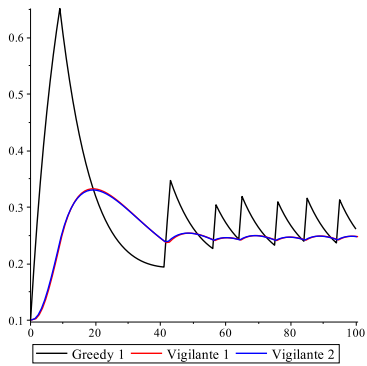}
\caption{Convergence in the case when $N=10$, $\lambda = 10$ and $\rho_1 = \rho_2 = 0.1$ and there are two vigilante players.}
\label{fig:TwoVigilante2}
\end{figure}

Finally, we consider the case with two vigilante players and one greedy player. As one would expect, the two vigilante players overestimate the greedy player's move and the system converges to a near deadlock state, with the two vigilante players unable to recover from the fact that they don't know about each other \cite{KKG13}. This is illustrated in Figure \ref{fig:TwoVigilante1}.
On the other hand, if the vigilantes adjust their $\rho_i$ $(i=1,2)$ upward to be more sensitive to their play, then the system does not converge, but oscillates as in the case with one greedy player and one vigilante player. In this case, however, the oscillation is about access rates $g$ that are almost fair. This is illustrated in Figure \ref{fig:TwoVigilante2}.

\section{Conclusions and Future Directions}\label{sec:Future}
In this paper, we formulated a multiplayer distributed resource access game in which some players have a greedy objective function and other players behave as vigilantes modifying their access probabilities to punish perceived greediness. Greedy players will back-off from a pure greedy strategy if the greedy strategy leads to poor payoff. We showed that the best response function for the greedy player under our formulation has a jump discontinuity, which leads to conditions under which there is no Nash equilibrium in the game. To understand this property, we formulated an exact representation for the greedy player's best response function in the case when there was one greedy player and one vigilante player. We used this formulation to show conditions under which a Nash equilibrium exists. We also illustrated that in the case when there is no Nash Equilibrium, then the discrete dynamic system generated from  fictitious play does not converge, but oscillates indefinitely as a result of the jump discontinuity. Finally, we discussed the cases when there was more than one greedy player and more than one vigilante.

In the future, we will investigate theoretical results on this model when there are a (small) number of vigilante and greedy players. It is clear from Figure \ref{fig:Oscillation} that the oscillations caused by the jump discontinuity have a somewhat complex periodic behavior. It would be interesting to understand how this periodicity is related to $\epsilon_g$ and $\epsilon_a$. In addition to this, we will study and compare in detail the discrete dynamical system arising from fictitious play to the continuous dynamics that arise from better-response dynamics (gradient descent or Jacobi iteration). Finally, there is a unique control theoretic problem embedded in this model. In the case where there were multiple vigilante's, we saw that it was easy for the vigilante's to overreact to each other. However, by modifying their respective $\rho_i$, the system was brought to a better point of (dynamic) stability (see Figures \ref{fig:TwoVigilante1} and \ref{fig:TwoVigilante2}). Dynamically controlling $\rho_i$ to improve system performance in the case of multiple greedy and vigilante players is of interest.

\section*{Acknowledgments}
Portions of Dr. Kesidis' work were supported by the National Science Foundation.

\bibliographystyle{IEEEtran}
\bibliography{mrabbrev,References}

\end{document}